# What is Science?

Pierre C Hohenberg,
New York University
December 2010

Abstract

This paper proposes a new definition of science based on the distinction between the activity of scientists and the product of that activity: the former is denoted (lower-case) science and the latter (upper-case) Science. These definitions are intended to clarify the nature of scientific knowledge, its authority as well as its limitations, and how scientific knowledge differs from other forms of human knowledge. The body of knowledge we call Science is exemplified by elementary arithmetic: it has the following properties: (i) Science is collective, public knowledge; (ii) Science is universal and free of contradiction; (iii) Science emerges from science; (iv) Science is nevertheless bathed in ignorance and subject to change. These properties imply that many questions that are of great interest to humanity are out of reach to Science, since they necessarily involve individual and group commitments and beliefs. Examples are questions of ethics, religion, politics, art and even technology, for which diversity is a fundamental virtue.

*Scientists need philosophers and historians of science like birds need ornithologists.*
(Attributed to Richard Feynman)

## I. Introduction

If you type the question in the above title into Google you will get an abundance of answers, many of them sensible and unsurprising. I will single out four of them for illustration:

    (i)   The National Academy of Sciences (2008)

*"The use of evidence to construct testable explanations and predictions of natural phenomena, as well as the knowledge generated through this process."*

    (ii)   Stephen Jay Gould (1997)

*The net of science covers the empirical universe: what is it made of (fact) and why does it work this way (theory).*

    (iii)   The US Supreme Court (1993) in *Daubert v. Merrell*

> *"Science is not an encyclopedic body of knowledge about the universe. Instead, it represents a process for proposing and refining theoretical explanations about*
> *the world that are subject to further testing and refinement. But, in order to qualify as 'scientific knowledge,' an inference or assertion must be derived by the scientific method. Proposed testimony must be supported by appropriate validation—i.e., 'good grounds,' based on what is known. In short, the requirement that an expert's testimony pertain to 'scientific knowledge' establishes a standard of evidentiary reliability."*

(iv) Richard Feynman (1968)

*Science alone of all the subjects contains within itself the lesson of the danger of belief in the infallibility of the greatest teachers in the preceding generation . . . As a matter of fact, I can also define science another way: Science is the belief in the ignorance of experts.*

The purpose of the present paper is first to refine the discussion by establishing a clear distinction between the human activity of scientists (hereafter "science" or "lower-case science") and the body of knowledge produced by that activity (which we will refer to as "Science" or "upper-case science"). We will attempt to describe and characterize the latter concept and to demonstrate why the distinction between science and Science leads to greater clarity. A second aim of this essay is to clarify the distinction between Science/science on the one hand and everything else, which we shall refer to as "non-Science".

What we call "science" is the human activity practiced by "scientists", who are often assembled in scientific communities, and whose results and pronouncements often conflict. The history of science is full of ups and downs, which have famously been likened to political revolutions. There are stated and unstated principles and rules by which scientists operate, primarily encoded in the "scientific method", which is in principle based on empirical observation, logic, verification and falsification, but in practice affected by many other contingencies.

Out of this scientific activity there emerges a body of knowledge we will call "Science", which has special properties that are quite different from those of "science" (the human activity) or of any other human activity. This body of knowledge, of which the clearest example is *arithmetic*, is a unique creation of the human community. It must be stressed, however, that the features of Science that lend it authority are also the source of *limitations* of Science, limitations that need to be identified and understood.

## II. The characteristics of Science

We will define what we mean by Science with reference to an idealized body of knowledge (e.g. arithmetic), to which all of science aspires, but which only reaches the status of Science by a mysterious process of *emergence*, over which no individual or group of individuals has direct control.

2.1 Science is collective, public knowledge

This knowledge belongs to no one and to everyone (to all mankind); it is anonymous. This is what we mean when we say that Science represents *objective* knowledge. We are not claiming that Science is not a product of human activity, but rather that it is one that has been liberated from its creator(s). Newton's Second Law ($F = ma$) is no more Newton's than *2+2=4* belongs to the person or persons who first conceived of arithmetic.

2.2 Science is universal

Scientific truths are intended to apply everywhere and are not tied to individuals or groups of individuals. There is no "German Science", "Jewish Science", "African Science". It follows that a new result will frequently *supersede* and invalidate an existing result, but this can only happen if it is *consistent* with previously established elements of Science. Any inconsistencies must be considered the unfinished business of Science, part of the sea of ignorance in which Science is bathed (see below).

The anonymity, universality and consistency of Science contrast sharply with other products of human endeavor, such as art, laws, religion and even technology. Beethoven's symphonies did not supersede Mozart's and they shall forever be associated with ("owned by") Beethoven. It is common for different nations to have different laws or customs and these differences are considered part of human diversity, not inconsistencies that must be eliminated. In contrast, well posed scientific questions have a unique answer.

2.3 How Science emerges from science

The mechanism by which this emergence occurs is shrouded in mystery and must be considered an empirical fact, rather than a fully understood process, but one thing is certain: it is not subject to individual control. What scientists do (even great ones like Newton and Einstein) is science, not Science. A piece of work is created by a scientist, who sends it out into the (scientific) world, where it becomes fair game for modification, criticism, verification and falsification. Most scientific work cause barely a ripple in the grand scheme of things, but in fortunate cases some works stimulate further activity or even become adopted as important, and eventually emerge (normally with modifications) into Science. Notable cases adorn the textbooks, often with more or less fanciful historical reconstructions. The important point is that once a result has emerged to the level of Science it is no longer "owned" by its originator.

2.4 Science is bathed in ignorance and subject to change

Scientific knowledge (Science), even though we characterize it as universal and objective, is nevertheless only partial and subject to change. It is much more stable than the knowledge produced directly by scientists (science) and is subject to a different dynamic, which causes it to change more slowly and smoothly than science. Apart from this lack of permanence, the knowledge inherent in Science is also *partial* in that it leaves many questions unanswered. The Nobel laureate David Gross has said (private communication) that "the most important product of science is ignorance", by which he is referring to new questions that are raised by important scientific results. Another expression of the same idea is the statement from H. Bondi (quoted in Morris, 2008, p. 7): "The power of science is the ability to say something without having to say everything".

The historian and philosopher of science Thomas Kuhn (1962) has famously used political revolutions and national conflicts as a metaphor for the development of science, but it is our view that a more apt metaphor, at least for Science, is provided by plate tectonics. Scientific knowledge is represented by land masses that emerge from an underwater world of scientific activity of individuals and scientific communities (science). All of this land is bathed by the sea and enveloped in a fog of humidity that constitute *ignorance*, i.e. a wealth of unanswered and perhaps unanswerable questions. As time progresses the land masses might rise, merge and evolve and the sea might well recede in places though there is no absolute guarantee that the sea level will always fall. The only mechanism for accretion of land mass is the activity of the (underwater) scientists who feed the continents (science), thus causing continual shifts in the shoreline between sea and land.

Another metaphor for Science, which emphasizes the requirement of consistency is the image of a *multiply connected web* (Anderson, 2001). Progress usually occurs on the periphery and it is difficult, though not impossible, to introduce new elements in the interior since so many links would have to be broken. This is the source of the (relative) stability as well as the authority of Science.

## IV. The paradigm of Science

It can be said that Science does not require belief: *2+2=4* whether you believe it or not. Nevertheless, the process of emergence of knowledge from science to Science is built on a generally accepted set of basic assumptions, a *paradigm* that seems to be implicit in the history of science (and of Science), at least over the past 400 years. This is not a credo that every scientist must adhere to, but rather a minimal set of implicit assumptions that the collective human phenomenon we are calling Science in practice rests on:

- The world of observable phenomena is real and intelligible in a collective manner.
- This public knowledge is subject to the requirements of logic and consistency.
- In addition to logic the knowledge must be based on observation and experimentation.

- Science is based on *naturalism*, by which we mean methodological naturalism. In particular *supernatural* explanations are rejected.

I am aware of the many philosophical issues surrounding naturalism, reductionism, materialism and the so-called mind-body problem, and my aim is certainly not to provide a solution of these problems, or even to add a substantive contribution to the philosophical debate. My aim, rather, is to identify the *minimal* assumptions that appear to be inherent in the pursuit and construction of Science. The term "methodological naturalism", seems to me to best summarize these minimal assumptions. Supernaturalism, on the other hand, allows for the intervention of causes affecting natural phenomena that are themselves not part of nature and might thus well contradict the laws of nature. This rejection of supernaturalism is an aspect of the requirement of consistency and not a detailed statement about the ontology of the world. In particular, I do not believe that physicalist reductionism, either of ontological or methodological variety, is a necessary presupposition of Science, though admittedly some prominent scientists would disagree with this view. Further discussion of these issues may be found in Hohenberg (2010).

## IV.     What is non-Science?

Apart from science, which we have already distinguished from Science, all other human activities belong to what we call "non-Science". This is the domain of art, literature, philosophy, politics, business, technology, medicine and of course, religion. The common characteristic of these diverse domains is the link to individuals or communities that must "own", "commit to" or "believe" the knowledge contained in the activity. Another characteristic is the celebration of diversity in these areas. What would be considered an inconsistency in Science is an expression of individual or collective freedom in art, politics or religion.

The strength of Science comes from the (intended) universality of its answers, but this is also a source of significant *limitations*, since many (some would say most) questions that interest people deeply are not subject to universal answers. Examples are:

- How do I find meaning in my life?
- Is this good or bad?
- How do I deal with the loss of a loved one?
- How do I face the prospect of my own death?
- Whom should I vote for?
- Does God exist?

It should be noted that many elements of non-Science involve large doses of Science or (science), for example medicine, technology, or even politics. Nevertheless, Science is properly used as a *tool* in these areas, to find appropriate answers in which, however, many other considerations, e.g. ethical or cultural values, must be brought to bear. These considerations necessarily engage the will or commitment of individuals or communities.

## V. What are alternatives to Science?

For questions that have or might have a universal answer the alternative to Science is not the various forms of non-Science, but rather *ignorance*, which in some measure is a component of each and every part of Science (see 2.4 above). For questions that have little or no expectation of possessing a universal answer the alternative, as we have said, is to be found in non-Science, for which Science is at most a useful tool but not an absolute source of authority. Such questions require belief, engagement, commitment, either individually or collectively. This distinction is inherent in our understanding of Science and it leads to crucial limitations of Scientific knowledge, which are often insufficiently stressed by enthusiasts of Science.

## VI. What are appropriate attitudes toward Science?

Since Science does not require belief and its tenets are not owned by any individual or group, it should be viewed as a *natural phenomenon*, akin to Niagara Falls, say, just not a material object. One does not argue with or contest Niagara Falls, one observes, admires or fears it and one might attempt to harness it for human benefit. In our view there are thus only three rational attitudes toward Science:

- You can ignore or distort Science (this is the option chosen by most people).
- You can try to learn Science.
- You can try to contribute to Science, but that activity will be what we call science. Whether it then emerges as Science is not under your control nor that of any individual, and if it eventually does emerge, it will no longer be associated with you as its originator.

In some real sense, Science is immune to criticism (like Niagara Falls). The main potential source of controversy in any scientific dispute is to decide if a statement or theory is indeed part of Science, as opposed to still being science, and if yes what the proper mix of knowledge and ignorance attaches to that statement. In terms of the geological analogy introduced in 2.4 above, the question is how far out of the water does this piece of land emerge and how much of the sea and fog of ignorance still envelops this land. In that sense, the distinctions and definitions we have introduced have not in themselves resolved any controversies. What we hope to have accomplished is to sharpen the questions in order to make correct answers more likely.

Note that our point of view is *not* that Science represents fact and non-Science represents value and thus Science is objective (see e.g. Harris, 2005). On the contrary, the values that Science is based on are what we have called the 'paradigm of Science' (Sec. IV) and these are a necessary component, along with the facts of Science. The difference is that both value and fact are *collective*, i.e. they must pass through the strong filter of universal or near-universal acceptance, which makes them both robust and restrictive, in that they exclude many questions and answers that cannot be (or have not been) collectivized. The operation of this filter is the process of emergence, which although always approximate in some sense, is unmistakable for true Science.

## VII. Science and religion

Elsewhere (Hohenberg, 2010) we have elaborated on the relationship between Science and religion, particularly as it affects the teaching of evolution, so we will only summarize that discussion here. As noted earlier, religion is the quintessential component of non-Science, addressing questions that have not led, and most likely never will lead, to universal answers. The age of universal religion essentially disappeared at the end of the Middle Ages, when it was replaced by the universal knowledge of Science, at least as an aspiration. Today, people who advocate universal religion are to be pitied or locked up, or both. The proper religious attitude toward Science is in my view either to ignore it as irrelevant to the aspirations of believers (appropriate in many cases), or to concentrate on those aspects which Science cannot confront (see IV above), taking care to remain consistent with Scientific knowledge. This is essentially the tactic used by Pope John Paul (1996) in his statement entitled "Truth Cannot Contradict Truth" concerning the theory of evolution. Religious attempts to confront Science on its own turf result at best in "junk science", which is the most one can say about Intelligent Design.

Science and religion do indeed represent "Nonoverlapping Magisteria" (NOMA) as stated by Gould (1997). What we have attempted to do is to explain more clearly why the magisteria do not overlap.

## VIII. Conclusion

Science is a body of knowledge that belongs to humanity as a whole. Its emergence and success over the past 400 years is a *miracle*. Science aims at, and thus attains at least approximately, objective and universal knowledge and it is thereby limited to those questions that can have universal answers. This contrasts with many other human endeavors (art, politics, religion) whose product is tied to individuals or communities and often requires an act of (implicit or explicit) volition, commitment or faith on the part of its adherents.

3